\documentclass[twoside,twocolumn,aps,prl,superscriptaddress,showpacs]{revtex4-1}
\usepackage[latin9]{inputenc}
\usepackage{bm}
\usepackage{amsmath}
\usepackage{amssymb}
\usepackage{graphicx}
\usepackage{esint}

\makeatletter
\@ifundefined{textcolor}{}
{%
 \definecolor{BLACK}{gray}{0}
 \definecolor{WHITE}{gray}{1}
 \definecolor{RED}{rgb}{1,0,0}
 \definecolor{GREEN}{rgb}{0,1,0}
 \definecolor{BLUE}{rgb}{0,0,1}
 \definecolor{CYAN}{cmyk}{1,0,0,0}
 \definecolor{MAGENTA}{cmyk}{0,1,0,0}
 \definecolor{YELLOW}{cmyk}{0,0,1,0}
}

\usepackage{bm}

\newcommand{\rl}{\rangle\!\langle}

\makeatother

\begin{document}

\title{Decoherence control by quantum decoherence itself}

\author{Katarzyna Roszak}
\email{katarzyna.roszak@pwr.wroc.pl}

\affiliation{Institute of Physics, Wroc{\l}aw University of Technology, 50-370
Wroc{\l}aw, Poland}

\author{Radim Filip}
\email{filip@optics.upol.cz}

\affiliation{Department of Optics, Palack\'{y} University, 17. listopadu 1192/12,
771 46 Olomouc, Czech Republic}

\author{Tom\'{a}\v{s} Novotn\'{y}}
\email{tno@karlov.mff.cuni.cz}

\affiliation{Department of Condensed Matter Physics, Faculty of Mathematics and
Physics, Charles University, 121 16 Prague, Czech Republic}

\date{\today}
\begin{abstract}
We propose a general approach of protecting a two-level system against
decoherence via quantum engineering of non-classical multiple superpositions
of coherent states in a non-Markovian reservoir. The scheme surprisingly
only uses the system-environment interaction responsible for the decoherence
and projective measurements of the two-level system. We demonstrate
the method on the example of an excitonic qubit in self-assembled
semiconductor quantum dots coupled to super-Ohmic reservoir of acoustic
phonons. 
\end{abstract}


\maketitle
\emph{Introduction}.--- Decoherence is the most significant obstacle
of expanding quantum technology. It appears as a result of an interaction
of the quantum system of our interest with an environment \cite{zurek03,*joos03,*schlosshauer07}.
The most common source of the decoherence is dephasing reducing a
quantum superposition between the eigen-states of energy of the system. If the environment, at least partially, resolves the
basis states of the system, their superposition is degraded or, ultimately,
it completely vanishes
\cite{imry97}. Frequently, the environment is not directly
controllable or measurable, it can be manipulated only by the same
interaction causing the decoherence which may represent a serious
limit. On the other hand, the system-environment interaction can produce
quantum entangled states between the system and the environment \cite{zurek82}.
The decoherence then becomes a quantum process which can be in principal
inverted, as opposed to the classical decoherence \cite{scully91,*haroche01}.
However, without a direct access to the environment the invertibility
is not feasible. Yet, quantum decoherence can still be used to {\em
pre-engineer} \cite{myatt00,*poyatos96} the environment to a state
which does not cause so destructive decoherence. 

As a very good practical example we can consider semiconductor quantum
dots (QDs), zero-dimensional nanostructures, in which charge carriers
display a discrete energy spectrum. A vast drawback for many applications
of semiconductor QDs is the carrier-phonon interaction which leads
to dephasing of electronic superpositions on picosecond timescales
\cite{borri01,*vagov04,vagov03}. To overcome this difficulty, a number
of solutions were proposed, including qubits coded on spin states
\cite{burkard99,*imamoglu99}, hybrid spin-charge schemes \cite{troiani03,*roszak05b},
modification of the optical-pulse shape \cite{hohenester04,*axt05a}
or reservoir properties \cite{cazayous04,*krummheuer05b}, and collective
encoding \cite{zanardi99a,*grodecka06}. Although some quite promising
results have been shown, a substantial reduction of decoherence is
accompanied by either amassing difficulties in coherent control of
the qubit (or many qubits), or by making the ensemble more involved
and resulting in fabrication problems. In this Letter, we propose
an inhibition of dephasing by reservoir pre-engineering assisted by the
same quantum dephasing process via repeated measurements of the qubit
state.

\emph{Quantum dephasing: toy model.}--- The simplest mechanism of
\emph{quantum dephasing} for a single energy-degenerate qubit can
be described by an interaction with a single environmental quantum
oscillator $E$ with vanishing frequency distinguishing between computational
basis states $|0\rangle$ and $|1\rangle$ of the qubit. The interaction
can be modeled by the interaction Hamiltonian $H_{I}=\kappa|1\rangle\langle1|P_{E}$,
where $\kappa$ is the interaction constant and $P_{E}=i(a_{E}^{\dagger}-a_{E})/\sqrt{2}$
is the momentum operator of the environmental oscillator (we use $\hbar=1$
throughout the paper). The interaction performs a non-demolition monitoring
of one of the degenerate states of the qubit, which does not change
the equal probabilities of the states $|0\rangle$ and $|1\rangle$
and only influences their superposition. In this case, the evolution
operator $U=|0\rangle\langle0|\otimes1_{E}+|1\rangle\langle1|\otimes U_{E}$
acting on both the qubit and the oscillator $E$ generates, if the
qubit is in the state $|1\rangle$, the unitary transformation of
the environmental states $U_{E}(\alpha)=\exp(i\kappa\tau P_{E})$,
with $\alpha=\kappa\tau/\sqrt{2}\in\mathbb{R}$, corresponding to
coherent displacement along the coordinate variable $X_{E}=(a_{E}+a_{E}^{\dagger})/\sqrt{2}$.
For the environmental oscillator being initially in the ground state
$|\mathrm{vac}\rangle_{E}$ ($\equiv\left|0\right\rangle _{E}$),
the unitary $U_{E}$ changes $|\mathrm{vac}\rangle_{E}$ to an overlapping
coherent state $|\alpha\rangle_{E}=U_{E}(\alpha)|\mathrm{vac}\rangle_{E}$.
If the qubit is initially in the superposition state $\big(|0\rangle+\exp(i\phi)|1\rangle\big)/\sqrt{2}$,
an entangled state $|\Psi_{0}\rangle=\big(|0\rangle|\mathrm{vac}\rangle_{E}+\exp(i\phi)|1\rangle|\alpha\rangle_{E}\big)/\sqrt{2}$
arises between the qubit and the environment. The square-root $D(\alpha)=|{}_{E}\langle\alpha|\mathrm{vac}\rangle_{E}|=\exp(-\alpha^{2}/2)$
of the overlap between the states of the environment then quantifies
both the amount of entanglement and phase damping process transferring
the initial qubit state to a mixture $\rho=\mathrm{Tr}_{E}\big(|\Psi_{0}\rangle\left\langle \Psi_{0}\right|\big)=\big(|0\rangle\langle0|+|1\rangle\langle1|+D(\alpha)\exp(i\phi)|1\rangle\langle0|+\mbox{h.c.}\big)/2$.

The generated entanglement by the dephasing can be exploited for state
preparation of the environment. Consider the qubit being prepared
initially in the state $|+\rangle=\big(|0\rangle+\exp(i\phi)|1\rangle\big)/\sqrt{2}$.
After it has undergone interaction with the environment for duration
$\tau$, the projection $|+\rangle\langle+|$ is executed on this
qubit \cite{nielsen00}. The environment $E$ is then projected to
the superposition state $|C_{1}(\alpha)\rangle_{E}=\big(|\mathrm{vac}\rangle_{E}+|\alpha\rangle_{E}\big)/\sqrt{2[1+D(\alpha)]}$.
The environment is thus engineered in a nonclassical quantum state
being a superposition of non-orthogonal states, known as the \emph{Schrödinger-cat
state} \cite{monroe96,*brune96}. To test, whether the superposition
state $|C_{1}(\alpha)\rangle_{E}$ present in the environment can
be better for a storage of qubit, a testing qubit only carrying information
in the phase variable $\phi$ is interacting for time interval $t$ with
the pre-engineered environment \emph{by the same type of interaction}
described by $H_{I}$. The resulting entangled state ($\beta\equiv\kappa t/\sqrt{2}\in\mathbb{R}$)
\begin{equation}
|\Psi_{1}\rangle=\frac{|0\rangle(|\mathrm{vac}\rangle_{E}+|\alpha\rangle_{E})+\exp(i\phi)|1\rangle(|\beta\rangle_{E}+|\alpha+\beta\rangle_{E})}{\sqrt{2[1+D(\alpha)]}},
\end{equation}
between the qubit and the environment is still subject to the quantum
dephasing. However, the overlap of $|C_{1}(\alpha)\rangle_{E}$ and
$U_{E}(\beta)|C_{1}(\alpha)\rangle_{E}$ is now substantially different
from $D(\alpha)$. Tracing out the environment, the qubit is then
described by the density matrix with the phase damping factor 
\begin{equation}
D_{1}(\alpha,\beta)=\frac{2D(\beta)+D(\alpha+\beta)+D(\alpha-\beta)}{2[1+D(\alpha)]}\label{gen}
\end{equation}
fully characterizing the dephasing process after engineering of the
environment. The last two terms arise due to interference effects
between the state preparation and the subsequent dephasing of the
testing qubit. If $\alpha=\beta$, then $D(\alpha-\beta)=1$ by definition.
On the other hand, since $D(\alpha)$, $D(\beta)$ and $D(\alpha+\beta)$
vanish for large $\alpha$ and $\beta$, the dephasing factor can
interestingly converge to $D_{1}=1/2$ for large equal interaction
times $\tau=t$. This should be contrasted with $D(\alpha\to\infty)=0$
for the initially ground state of the environment.

This is a remarkable result, since by a conditional engineering of
the environment using the same quantum dephasing process we are able
to protect the subsequent qubit evolution against the very same dephasing
mechanism. The protection arises due to a quantum interference term
$D(\alpha-\beta)$ in Eq.~\eqref{gen} caused by the principal indistinguishability
of the state $|\alpha\rangle_{E}$ being a component in both the states
$\big(|\mathrm{vac}\rangle_{E}+|\alpha\rangle_{E})\big)/\sqrt{2[1+D(\alpha)]}$
and $\big(|\alpha\rangle_{E}+|2\alpha\rangle_{E}\big)/\sqrt{2[1+D(\alpha)]}$
induced by the dephasing interaction for $\tau=t$ in the environment.
Is the superposition in the environment really required? Imagine that
the engineered superposition collapses into the incoherent mixture
$\big(|\mathrm{vac}\rangle_{E}\langle\mathrm{vac}|_{E}+|\alpha\rangle_{E}\langle\alpha|_{E}\big)/2$
before the testing qubit is interacting with the environment. The
dephasing factor then remains $D_{1}^{\mathrm{inc}}(\alpha,\beta)=\exp(-\beta^{2}/2)$,
the same as without any environment engineering. Therefore, the quantum
superposition of (non-orthogonal coherent) environmental states becomes
a \emph{resource} necessary for our method of protecting qubits. Quantum
dephasing therefore has the principal feature which allows to be corrected
by itself, differently from the classical dephasing.

For the initially ground state of the environment, after $M$ identical
repetitions of the state preparation with preparation times $\tau$,
the state superposition $|C_{M}(\alpha)\rangle_{E}=\sum_{k=0}^{M}{M \choose k}|k\alpha\rangle_{E}/\sqrt{N_{M}}$
of the environmental coherent states is generated. This special state,
a superposition of equidistantly displaced states with the coefficients
proportional to the combinatorial numbers from the Pascal triangle,
is a direct outcome of the \emph{quantum random walk} with coherent
states in the environment and yields for the decoherence factor $D_{M}(\alpha)\equiv|_{E}\langle C_{M}(\alpha)|U_{E}(\alpha)|C_{M}(\alpha)\rangle_{E}|$
the expression 
\begin{equation}
D_{M}(\alpha)=\frac{\sum_{k,l=0}^{M}{M \choose k}{M \choose l}\exp\left(-(k-l-1)^{2}\frac{\alpha^{2}}{2}\right)}{\sum_{k,l=0}^{M}{M \choose k}{M \choose l}\exp\left(-(k-l)^{2}\frac{\alpha^{2}}{2}\right)}.\label{eq:DM}
\end{equation}
For small $\alpha\ll1$ and large $M\gg1$, due to overlaps of the
states $|k\alpha\rangle_{E}$ the state $|C_{M}(\alpha)\rangle_{E}$
approaches a pure Gaussian state squeezed in the momentum variable
$P_{E}$ with the variance of the momentum $\langle(\Delta P_{E})^{2}\rangle=1/[2(1+\alpha^{2}M/2)]$
calculated in the Supplemental Material. Consequently, $D_{M}(\alpha\ll1)=|\langle\exp(i\sqrt{2}\alpha P_{E})\rangle_{C_{M}(\alpha)}|=\exp\left(-\alpha^{2}\langle(\Delta P_{E})^{2}\rangle\right)=\exp\left[-1/\left(2/\alpha^{2}+M\right)\right]$,
which is increasing with $M$. The measurement-induced squeezing of
the reservoir momentum $P_{E}$ explains why the interaction $H_{I}=\kappa|1\rangle\langle1|P_{E}$
is less dephasing the qubit, since the variable $P_{E}$ is less fluctuating.
As shown numerically in the Supplemental Material the above formula
approximates very well Eq.\ \eqref{eq:DM} even for large $\alpha$'s
and we find the asymptotic behavior for sufficiently large $M\gg\max\{1,\alpha^{-2}\}$
\begin{equation}
D_{M}(\alpha)=1-\frac{1}{M}+\mathcal{O}\left(\frac{1}{M^{2}}\right).\label{eq:DMasympt}
\end{equation}

This result implies that the dephasing process can be completely
stopped by the repeated state engineering based on the system-environment
interaction which is itself responsible for the dephasing. However,
it is unclear whether properties of this simplistic case carry over
to more realistic situations involving nondegenerate qubits and environments
with a large number of finite frequency modes. As we show in detail
below, the answer is positive and we identify a whole class of experimentally-relevant
solid-state setups where an analogous mechanism of decoherence suppression
can be implemented.

\emph{Infinite reservoir model \& its free dynamics}.--- The system
under study consists of a self-assembled, single level quantum dot
under the influence of a reservoir of longitudinal acoustic phonons
described by $H_{\mathrm{ph}}=\sum_{\bm{k}}\omega_{\bm{k}}b_{\bm{k}}^{\dag}b_{\bm{k}}$,
with $\omega_{\bm{k}}=vk$ being the frequency of the phonon mode
with the wave vector $\bm{k}$ ($v$ is the speed of longitudinal
sound waves). We consider just two electronic states of the dot forming
the qubit: $|0\rangle$ when the dot is in its ground state (``empty'',
i.e.\ no exciton) and $|1\rangle$ indicating the excited QD (``occupied''
with an exciton in its ground state) with bare excitation energy $\tilde{\epsilon}$,
i.e., $H_{{\rm dot}}=\tilde{\epsilon}|1\rl1|$. When occupied by the
exciton, the dot experiences interaction with the phonon environment
by means of the deformation potential coupling \cite{grodecka05a,mahan00}
$H_{{\rm int}}=|1\rl1|\sum_{\bm{k}}(f_{\bm{k}}^{*}b_{\bm{k}}+f_{\bm{k}}b_{\bm{k}}^{\dag})$
with the super-Ohmic spectral density $J(\omega>0)=\sum_{\bm{k}}|f_{\bm{k}}|^{2}\delta(\omega-\omega_{\bm{k}})=\eta\omega^{3}e^{-(\omega/\omega_{c})^{2}}F(\omega/\omega_{c})$
characterized by the low-frequency coefficient $\eta\doteq0.027\,\mathrm{ps^{2}}$,
size-dependent high-frequency cut-off $\omega_{c}\doteq7.21\,\mathrm{ps^{-1}}$,
and ``form-factor'' $F(x\ll1)\approx1,\, F(x\gg1)\approx1/(48x^{2})$
corresponding to the typical material and spatial parameters for a
self-assembled InAs/GaAs structure found in Ref.~\cite{roszak10}
with anisotropic Gaussian exciton wave functions of $5$ nm width
in the $xy$ plane and $1$ nm along $z$ (for details see the Supplemental
Material). The exciton-phonon interaction term in the Hamiltonian
is linear in phonon operators and describes a shift of the lattice
equilibrium induced by the presence of a charge distribution in the
dot associated with the classical energy of the displaced oscillators
$\varepsilon_{\mathrm{cl}}=\sum_{\bm{k}}|f_{\bm{k}}|^{2}/\omega_{\bm{k}}\equiv\int_{0}^{\infty}d\omega J(\omega)/\omega$.
The total Hamiltonian $H=H_{{\rm dot}}+H_{\mathrm{ph}}+H_{{\rm int}}$
being a variant of exactly-solvable independent boson models is diagonalized
\cite[Sec.\ 4.3.1]{mahan00} by a canonical transformation represented
by the unitary operator $S=\exp\big[|1\rl1|\sum_{\bm{k}}(f_{\bm{k}}b_{\bm{k}}^{\dag}-f_{\bm{k}}^{*}b_{\bm{k}})/{\omega_{\bm{k}}}\big]\equiv\exp(-i|1\rl1|B)$
yielding $SHS^{\dagger}=\epsilon|1\rl1|+H_{\mathrm{ph}}$, with renormalized
(physical) exciton energy $\epsilon=\tilde{\epsilon}-\varepsilon_{\mathrm{cl}}$
taken equal to 1 eV.

Dynamics of the quantum dot represented by its reduced density matrix
$\rho_{ij}(t)=\langle i|\mathrm{Tr}_{\mathrm{ph}}\big[e^{-iHt}\sigma(0)e^{iHt}\big]|j\rangle,\ i,j=0,1$
can be solved exactly for factorizing initial conditions $\sigma(0)=\rho(0)\otimes\varrho_{\mathrm{ph}}^{\mathrm{can}}$
with canonical state of the phonon reservoir $\varrho_{\mathrm{ph}}^{\mathrm{can}}=e^{-\beta H_{\mathrm{ph}}}/\mathrm{Tr}_{\mathrm{ph}}\big(e^{-\beta H_{\mathrm{ph}}}\big)$
at inverse temperature $\beta=1/k_{B}T$. Diagonal elements are constant
$\rho_{00}(t)=\rho_{00}(0),\ \rho_{11}(t)=\rho_{11}(0)$, i.e., there
is no phonon-induced exciton relaxation, while the time evolution
of the off-diagonal elements $\rho_{01}(t)=\rho_{10}^{*}(t)$ describing
the decoherence of superposition states between $|0\rangle$ and $|1\rangle$
exhibits non-exponential, i.e., non-Markovian decay $\rho_{01}(t)/(e^{i\epsilon t}\rho_{01}(0))=\mathrm{Tr}_{\mathrm{ph}}[\varrho_{\mathrm{ph}}^{\mathrm{can}}e^{iB(-t)}e^{-iB(0)}]\equiv\langle W(t)\rangle_{0}$
with the Weyl operator $W(t)\equiv e^{iB(-t)}e^{-iB(0)}$ \cite{roszak06a,roszak06b}.
Its equilibrium mean value $\langle W(t)\rangle_{0}\equiv\exp[-w(t)]=\exp\left[\langle B(-t)B(0)\rangle_{0}-\langle B(0)^{2}\rangle_{0}\right]$
\cite{weiss99,mahan00} is governed by the bath correlation function
\begin{equation}
w(t)=\int_{0}^{\infty}d\omega\frac{J(\omega)}{\omega^{2}}\left[(1-\cos\omega t)\coth\frac{\beta\omega}{2}-i\sin\omega t\right].
\end{equation}

The model thus shows features of \emph{pure dephasing}, i.e., only
the coherences, which can be measured by the amplitude of coherent
dipole radiation emitted by the dot, decay with time. Moreover, for
the super-Ohmic spectral density characteristic of this system, due
to the Riemann-Lebesgue lemma the decay saturates at a finite value
$w(|t|\gg\tau_{\phi})\equiv w_{\infty}=\int_{0}^{\infty}d\omega\tfrac{J(\omega)}{\omega^{2}}\coth\tfrac{\beta\omega}{2}$
for times much longer than the dephasing time $\tau_{\phi}\sim\mathrm{min}(1/\omega_{c},\hbar/k_{B}T)$,
thus the pure dephasing is only \emph{partial} or \emph{incomplete}
\cite{roszak06a,Hornberger:book09,*Ropke:CMP12}. In the zero-temperature
limit $(\beta\to\infty)$ the asymptotic value of the coherence reads
$D\equiv|\rho_{01}(t\gg\tau_{\phi})|/|\rho_{01}(0)|\equiv \exp(-w_{\infty})=\exp[-\int_{0}^{\infty}d\omega J(\omega)/\omega^{2}]=|\langle\mathrm{vac}|e^{-iB}|\mathrm{vac}\rangle|^{2}=|\langle\mathrm{vac}|\widetilde{\mathrm{vac}}\rangle|^{2}$,
where $|\mathrm{vac}\rangle,\,|\widetilde{\mathrm{vac}}\rangle$ are
the phonon vacua when the QD is empty or occupied, respectively. The
overlap of the two mutually displaced vacua is non-zero, which means
that despite of the continuous spectrum of phonon modes the orthogonality
catastrophe is incomplete --- this reflects the asymptotic nature
of the couplings $f_{\bm{k}}$ for small $\bm{k}$'s (and $\omega$)
due to identical phonon coupling to electrons and holes for long phonon
wavelength \cite{krummheuer02} resulting in the super-Ohmic spectral
density of exciton-phonon coupling. Consequently, for small $\bm{k}$'s
the trace left by the exciton in the bath is too weak to be distinguished
from the vacuum case and, thus, decoherence is only partial \cite[Sec.~3]{imry97}.

\emph{Repeated initializations}.--- We may study not only the state
of the QD considered so far but also the state of the phononic subsystem
analogously to the above toy model.
The creation of an exciton in the QD perturbs the phonon reservoir
state by shifting the coordinates. If the exciton is created in a
superposition state, the phonon reservoir will react by following
in parallel two different evolutions coherently superposed. Now, we
may ask again what is the effect of repeated measurements of the dot
state on the degree of the partial pure dephasing. Therefore, we analyze
the evolution of the composite system of the dot and the phonon reservoir
subject to strong projective measurements \cite{nielsen00} performed
on the QD subsystem. Each measurement is represented by orthonormal
projection operators of the form $P_{\pm}=|\pm\rangle\langle\pm|\otimes\mathbb{I}$
with complementary and orthonormal pure qubit states $|\pm\rangle=(|0\rangle\pm e^{i\phi}|1\rangle)/\sqrt{2}$
and the unity in the reservoir subsystem $\mathbb{I}$.
We consider free evolution of the composite system starting from a
factorized initial/reinitialized condition $\sigma_{\mathrm{init}}=|\mathrm{init}\rangle\langle\mathrm{init}|\otimes\varrho_{\mathrm{init}}$
corresponding either to the true initial condition or to an output
of previous measurement (see Eq.~\eqref{measurement} below) with
the initial state of the QD qubit $|\mathrm{init}\rangle=(|0\rangle+e^{i\phi_{\mathrm{init}}}|1\rangle)/\sqrt{2}$
and an arbitrary phonon reservoir density matrix $\varrho_{\mathrm{init}}$.
We choose the equal-weight superpositions so that neither the dephasing
interaction nor the measurement processes regardless of their outcome
change the occupation factors and only influence the coherences.

Under these assumptions the state of the composite system right after
the measurement at time $\tau$ with the outcome $\pm$ is given by
\begin{equation}
\sigma_{\pm}(\tau^{+})=\frac{P_{\pm}\sigma(\tau)P_{\pm}}{\mathrm{Tr}_{\mathrm{ph}}\left(\langle\pm|\sigma(\tau)|\pm\rangle\right)}=|\pm\rangle\langle\pm|\otimes\varrho_{\pm}(\tau^{+}),\label{measurement}
\end{equation}
with the measurement-outcome-dependent phonon reservoir density matrices\begin{widetext}
\begin{equation}
\varrho_{\pm}(\tau^{+})=\frac{\varrho_{\mathrm{init}}(\tau)+W^{\dagger}(\tau)\varrho_{\mathrm{init}}(\tau)W(\tau)\pm\left[e^{i\Delta\phi(\tau)}\varrho_{\mathrm{init}}(\tau)W(\tau)+h.c.\right]}{2\left(1\pm\Re\left[e^{i\Delta\phi(\tau)}\langle W(\tau)\rangle_{\varrho_{\mathrm{init}}(\tau)}\right]\right)}.\label{densitymatrices}
\end{equation}
\end{widetext}Here, $\Re$ denotes the real part, $h.c.$ the hermitian
conjugate, $\varrho_{\mathrm{init}}(\tau)=e^{-iH_{\mathrm{ph}}\tau}\varrho_{\mathrm{init}}e^{iH_{\mathrm{ph}}\tau}$,
$\Delta\phi(\tau)\equiv\epsilon\tau+\phi-\phi_{\mathrm{init}}$, and
$\langle W(\tau)\rangle_{\varrho_{\mathrm{init}}(\tau)}$ denotes
the average of the Weyl operator with respect to this time-evolved
phonon density matrix. The respective measurement outcomes are
obtained with probabilities $p_{\pm}(\tau)=\mathrm{Tr}_{\mathrm{ph}}\left(\langle\pm|\sigma(\tau)|\pm\rangle\right)=\left\{ 1\pm\Re\left[e^{i\Delta\phi(\tau)}\langle W(\tau)\rangle_{\varrho_{\mathrm{init}}(\tau)}\right]\right\} /2$.

Note furthermore, that regardless of the measurement outcome, the
degree of coherence just after the measurement is fully restored to
unity $D_{1}(\tau,0^{+})=|\langle0|\rho(\tau^{+})|1\rangle|/|\langle0|\rho_{\mathrm{init}}|1\rangle|=1$,
i.e., the net outcome of the measurement on the state of the qubit
is, apart from a possible (controlled) phase shift, just the reinitialization
of the qubit state (compare with Eq.~\eqref{gen} for $\beta\to0^{+}$).
However, the state of the phonon reservoir does change and this has
important consequences for further evolution of the qubit. The scheme
outlined above can be iterated to yield results for an arbitrary series
of measurements, but it acquires great complexity rapidly with
the growing number of measurements. It is therefore convenient to
study just the single-measurement scenario, especially since an observable decrease of
dephasing can be detected already there.

\emph{Single measurement case}.--- In the following, we study time
evolution of the qubit at time $t$ after the measurement performed
time $\tau$ after its initialization, in particular we monitor the
degree of coherence $D_{1}^{\pm}(\tau,t)=|\rho_{01}^{\pm}(t+\tau)|/|\rho_{01}(0)|=|\rho_{01}^{\pm}(t+\tau)|/|\rho_{01}^{\pm}(\tau^{+})|$
as functions of the delay time $t$, measurement time $\tau$ and
the measurement outcome ($\pm$). To this end we evolve the density
matrices from Eq.~\eqref{measurement} for the time span $t$ and
then evaluate the coherences $\rho_{01}^{\pm}(t+\tau)$. Calculation
follows the line analogous to the free evolution discussed above with
the initial thermal density matrix $\varrho_{\mathrm{ph}}^{\mathrm{can}}$
replaced with those of Eq.~\eqref{densitymatrices} leading to $\rho_{01}^{\pm}(t+\tau)/(e^{i\epsilon t}\rho_{01}^{\pm}(\tau^{+}))=\langle W(t)\rangle_{\varrho_{\pm}(t+\tau)}$.
Using the fact that with $\varrho_{\mathrm{init}}=\varrho_{\mathrm{ph}}^{\mathrm{can}}$
we also get $\varrho_{\mathrm{init}}(\tau)=\varrho_{\mathrm{ph}}^{\mathrm{can}}$,
the result reads (for details see the Supplemental Material)\begin{widetext}
\begin{equation}
D_{1}^{\pm}(\tau,t)=\left|e^{-w(t)}\frac{1+e^{w(t)-w(-t)+w(\tau)-w(-\tau)-w(t+\tau)+w(-t-\tau)}\pm e^{i\Delta\phi(\tau)}e^{-w(t+\tau)+w(t)}\pm e^{-i\Delta\phi(\tau)}e^{-w(-t)-2w(-\tau)+w(-t-\tau)}}{2(1\pm\Re[e^{i\Delta\phi(\tau)}e^{-w(\tau)}])}\right|.\label{eq:D1}
\end{equation}
This result is proportional to $D(t)\equiv\left|e^{-w(t)}\right|=e^{-\Re\left[w(t)\right]}$
which means that the asymptotic value for large times $t$ can only
be nozero if $D(t\to\infty)\equiv \exp(-w_{\infty})\equiv D>0$, i.e.,
for partial dephasing, as is the case for the super-Ohmic bath. We
then get for large times $t,\tau\gg\tau_{\phi}$ (we use $\phi\equiv\Delta\phi(\tau)$) 
\begin{equation}
D_{1}^{\pm}(\phi)=D\frac{\left|2\pm e^{i\phi}\pm D{}^{2}e^{-i\phi}\right|}{2(1\pm D\cos\phi)}=D\frac{\sqrt{5+D^{4}\pm4(1+D^{2})\cos\phi+2D^{2}\cos2\phi}}{2(1\pm D\cos\phi)}.\label{das}
\end{equation}

\end{widetext}

\begin{figure}[th]
\begin{centering}
\includegraphics[width=1\columnwidth]{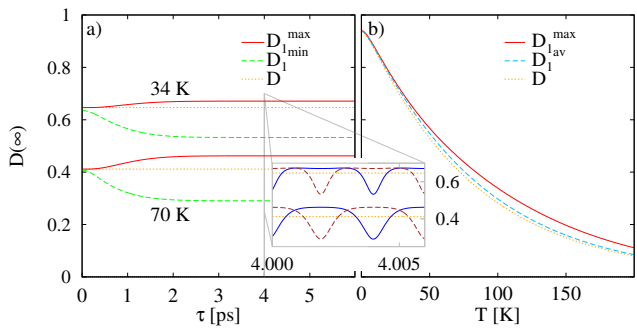} 
\par\end{centering}

\caption{(Color online) a) Asymptotic degree of coherence as a function of
the delay time for two temperatures. The envelopes of the maximal
(solid red line) and minimal (dashed green line) values of $D_{1}^{\pm}(\tau,t\to\infty)$
\eqref{eq:D1} are shown together with the detailed time evolution
for the measurement outcome $|+\rangle$ (solid blue line) and $|-\rangle$
(dashed brown line) on a much shorter time scale in the inset. b)
Maximal value (full red line) as well as the averaged one (dashed
cyan line; see the main text for details) of the asymptotic degree
of coherence for a range of temperatures. In both panels, the dotted
orange lines denote the degree of coherence in the  measurement-free case.\label{fig:asodtau}}
\end{figure}

Obviously, these values oscillate as functions of the delay time $\tau$
between the preparation of the qubit and its measurement with the
frequency determined by the shifted exciton energy $\epsilon$ (corresponding
period is on the order of few femtoseconds) as depicted in the inset
of Fig.~\ref{fig:asodtau}. We also plot there the envelopes of curves
\eqref{das} on the longer timescale of picoseconds showing the saturation
of the initial sub-picosecond transient behavior. The overall magnitude
of the asymptotic degree of coherence decreases with increasing temperature
as presented in Fig.~\ref{fig:asodtau}b). Let us now analyze the
formulas \eqref{das} in more detail. First, $D^{-}$ is easily obtained
from $D^{+}$ by the phase shift $\phi\rightarrow\phi+\pi$ so that
it suffices to study the latter one. It always attains a minimum $D(1+D)/2$
at $\phi=\pi$ and has a local extremum $D(3+D^{2})/[2(1+D)]$ at $\phi=0$.
For small enough $D\leq D_{c}\doteq0.48$ ($D_{c}\in[0,1]$ is determined
by $D_{c}^{3}+2D_{c}^{2}+3D_{c}-2=0$) this extremum is the global
maximum, while for larger $D\geq D_{c}$ it is just a local minimum
and the maximum $(1+D)\sqrt{D(1+D)/(D^{2}+3D+4})$ is realized at
$\phi_{\mathrm{max}}=\arccos\left[(2-D-2D^{2}-D^{3})/2D\right]$.
The difference of the maximal value from the free case value
$D$ is maximized for $D_{\mathrm{max}}^{\mathrm{max}}=\left(\sqrt{17}-3\right)/4\doteq0.28$
(corresponding to $T\approx120$ K) by the excess value of $\left(71-17\sqrt{17}\right)/16\doteq0.057$,
some 20\% above the free case. As discussed in more detail below we
may be also interested in the weighted average $D_{1}^{\mathrm{av}}\equiv\intop_{0}^{2\pi}\tfrac{d\phi}{2\pi}[p_{+}(\phi)D_{1}^{+}(\phi)+p_{-}(\phi)D_{1}^{-}(\phi)]$
which is bigger than $D$ since the integrand is never below $D$
(equality happens only at $\phi=0,\pi$). Numerical analysis reveals
that the maximum difference from the free case is obtained at $D_{\mathrm{max}}^{\mathrm{av}}\doteq0.47$
(corresponding to $T\approx60$ K) with the magnitude roughly $0.019$,
about 4\% of the free case value. These conclusions are consistent
with the plots in Fig.~\ref{fig:asodtau}b).

\emph{Experimental feasibility}.--- We have analyzed thus far properties
of an idealized model and it is necessary to scrutinize whether our
conclusions can be carried over to the experimentally realistic situations.
There are several points which might in principle endanger our conclusions.
First, we have only considered the Hamiltonian describing the free
evolution, which is purely harmonic in the acoustic phonon modes and
the excitonic interaction with them is solely of pure dephasing type.
In reality there are also optical phonons which cause the relaxation
of the exciton occupation and, moreover, there is radiative relaxation
channel too --- these effects, however, become effective only at much
longer timescales on the order of tens or hundreds picoseconds \cite{borri01}
while our asymptotic times are just a few picoseconds. Since the dephasing-suppression
mechanism hinges on the creation of ``cat states'' of the acoustic
reservoir modes, their potential dephasing beyond the excitonic interaction
by anharmonic terms or by coupling to other (e.g., optical) modes
would be detrimental to the predicted effect. While such effects do
exist and may be relevant in certain contexts (see, e.g., Ref.~\cite{machnikowski06d}),
the estimated lifetime of the acoustic phonons \cite{hu11} is on
the order of 1 nanosecond, which makes these issues irrelevant for
our discussion. Finally, we have assumed an instantaneous projective
measurement of the qubit state. However, existing
projective measurements are achieved by optical pulses whose duration
is at least ten(s) femtoseconds during which the freely evolving qubit
phase $\epsilon\tau$ acquires several multiples of $2\pi$'s (see
the inset of Fig.~\ref{fig:asodtau}). Thus, one might expect that
the effect would be smeared by the phase averaging. Nevertheless, finite
duration of pulses is not necessarily fatal to possible proof-of-principle tests. What
matters is the short duration of the pulse with respect to the characteristic
time scale of the \emph{phonons }being on the order of 1 ps ($\approx1/\omega_{c}$)
and the ability to very precisely control the relative phase between
the initialization and measurement pulses. This is currently possible
by splitting the initial pulse and using the optical delay line with
exquisite sub-cycle tuning of the relative phase as realized in pump-probe
and multidimensional optical spectroscopies \cite{mukamel95,borri01}.
Thus, the approximation of delta-like pulses is done and justified
for the study of phonon dynamics \cite{vagov03,mukamel95}. Even if the experiment
is not completely controlled (the relative phase $\epsilon\tau$ is
fluctuating between subsequent runs of the measurement) and/or the
measurement outcomes of the qubit state are ignored (e.g., to avoid
discarding data), the averaged result described by the quantity $D_{1}^{\mathrm{av}}$
introduced above and plotted in Fig.~\ref{fig:asodtau}b) still shows
enhancement over the free case, although its absolute magnitude is 3-times
less than in the fully controlled case. Altogether,
we believe that the predicted effect should be experimentally observable.

\emph{Conclusions}.--- We have proposed measurement-induced quantum
pre-engineering of non-Markovian environment consisting of a super-Ohmic
reservoir of longitudinal acoustic phonons which can be directly exploited
to control quantum-dot-based qubit decoherence using only the single
type of coupling between the qubit and the environment. A feasible
proof-of-principle experimental test of the proposed method with self-assembled
semiconductor quantum dots would be a practical test of the quantum
nature of dephasing for a solid state system. The method can be also translated to the cavity QED, atomic, or trapped ion experiments.

\textbf{Acknowledgments} This work was financially supported by the
Polish NCN Grant No.~2012/05/B/ST3/02875 (K.R.~and T.N.), by the
TEAM programme of the Foundation for Polish Science, co-financed from
the European Regional Development Fund (K.R.) and the Czech Science
Foundation via projects No.~P205/12/0577 (R.F.) and P204/12/0853
(T.N.).

\section*{Supplemental Material}

\subsection{Environment engineering by many repetitions (derivation of Eq.~\eqref{eq:DMasympt})}

After the $M$-times identical state preparation with the preparation
times $\tau$ and subsequent evolution of the qubit during the same
time $\tau$, the phase damping factor is defined as an absolute value
of the scalar product between states 
\begin{equation}
|C_{M}(\alpha)\rangle_{E}=\frac{1}{\sqrt{N_{M}}}\sum_{k=0}^{M}{M \choose k}|k\alpha\rangle_{E},\label{state1}
\end{equation}
and the displaced version 
\begin{equation}
U_{E}(\alpha)|C_{M}(\alpha)\rangle_{E}=\frac{1}{\sqrt{N_{M}}}\sum_{l=0}^{M}{M \choose l}|l\alpha+\alpha\rangle_{E},\label{state2}
\end{equation}
with the same normalization factor $N_{M}=\sum_{k,l=0}^{M}{M \choose k}{M \choose l}\exp\left(-(k-l)^{2}\alpha^{2}/2\right)$. 

\begin{figure}
\includegraphics[width=1\columnwidth]{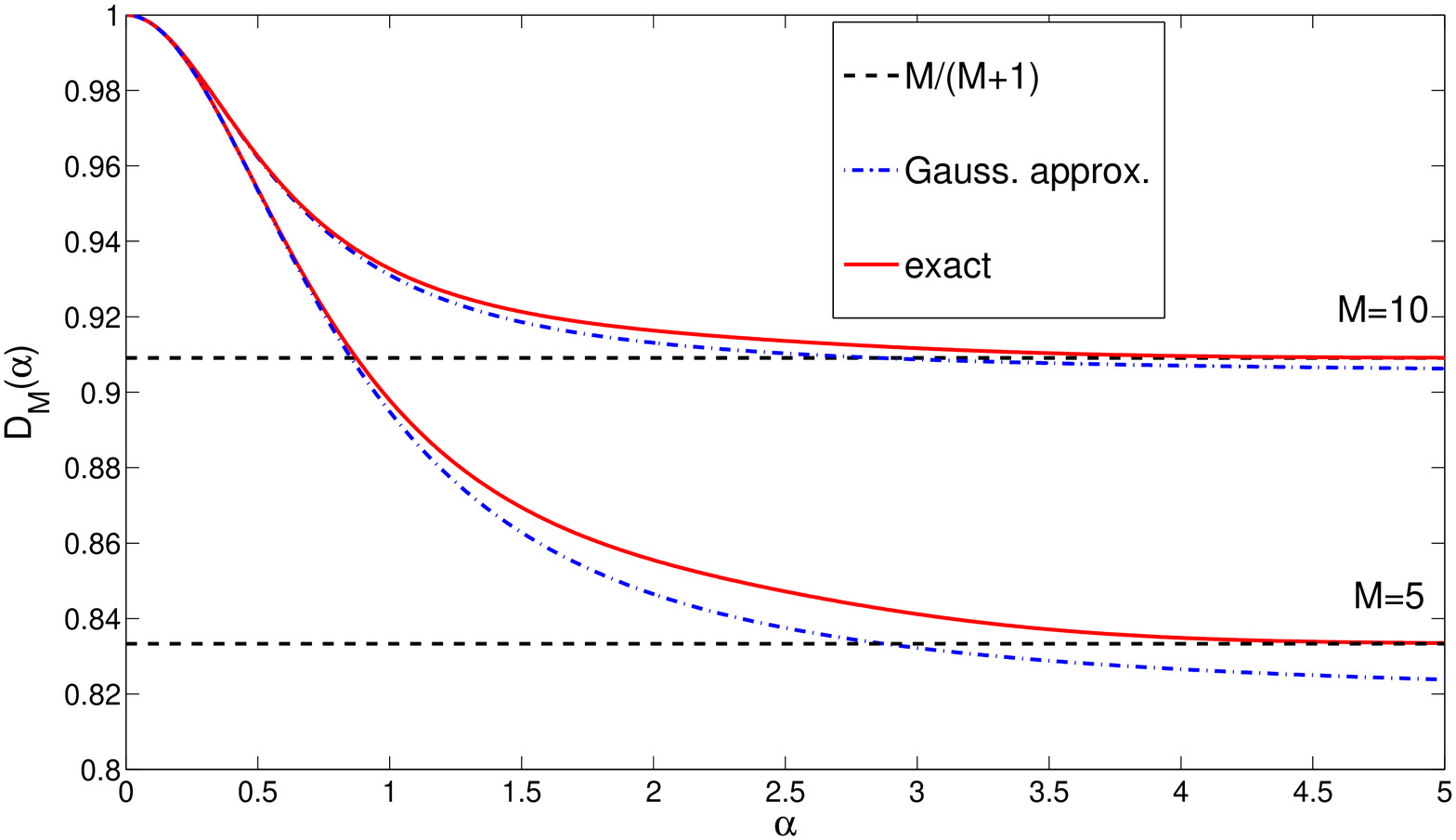}\caption{(Color online) Asymptotic decoherence factor $D_{M}(\alpha)$ for two values of $M=5$
(lower set of curves) and $M=10$ (upper curves) as functions of the
integrated interaction strength $\alpha$. Exact expression \eqref{eq:DM}
(full red lines) is compared with the Gaussian approximation \eqref{eq:Gauss}
(blue dash-dotted lines) and the asymptotic value \eqref{eq:DMinf}
(black dashed lines). \label{fig:Asymptotic-decoherence-factor}}
\end{figure}

The coherent states $|k\alpha\rangle_{E}$ with vanishing mean of
momentum $P_{E}$ can be expressed in the coordinate representation
of the operator $X_{E}=(a_{E}+a_{E}^{\dagger})/\sqrt{2}$ in the form
\begin{equation}
\langle x|k\alpha\rangle_{E}=\frac{1}{\pi^{\frac{1}{4}}}\exp\left(-\frac{(x-\sqrt{2}k\alpha)^{2}}{2}\right).
\end{equation}
In the limit of small $\alpha\ll1$ the constituents of the sums \eqref{state1}
and \eqref{state2} highly overlap and form smooth resulting wave
functions. Moreover, at large $M\gg1$ we can approximate the binomial
coefficients by expansion based on the Stirling formula ${M \choose k}\equiv{M \choose M/2+\xi\sqrt{M/2}}\approx2^{M}e^{-\xi^{2}}$
yielding for the normalization factor 
\begin{equation}
\begin{split}N_{M} & \approx2^{2M-1}M\int_{-\infty}^{\infty}d\xi\int_{-\infty}^{\infty}d\eta e^{-\xi^{2}-\eta^{2}-\frac{\alpha^{2}M}{4}(\xi-\eta)^{2}}\\
 & =\frac{2^{2M-1}M\pi}{\sqrt{1+\frac{\alpha^{2}M}{2}}}
\end{split}
\end{equation}
 and, similarly, for the whole wave function 
\begin{equation}
\langle x|C_{M}(\alpha)\rangle_{E}\approx\frac{1}{\sqrt[4]{\pi\left(1+\frac{\alpha^{2}M}{2}\right)}}e^{-\frac{\left(x-\langle X_{E}\rangle\right)^{2}}{2\left(1+\frac{\alpha^{2}M}{2}\right)}}.
\end{equation}
It is a pure Gaussian state in the environment with the mean $\langle X_{E}\rangle=\alpha M/\sqrt{2}$
and variance $\langle\left(\Delta X_{E}\right)^{2}\rangle=\left(1+\alpha^{2}M/2\right)/2$.
Consequently, the variance in the momentum reads ($\langle P_{E}\rangle=0$)
\begin{equation}
\langle\left(\Delta P_{E}\right)^{2}\rangle=\frac{1}{2}\left(1+\frac{\alpha^{2}M}{2}\right)^{-1}
\end{equation}
as stated in the main text. This description based on the pure Gaussian
state in the environment giving 
\begin{equation}
D_{M}^{\mathrm{Gauss}}(\alpha)=e^{-\frac{1}{\frac{2}{\alpha^{2}}+M}}\label{eq:Gauss}
\end{equation}
 very satisfactorily approximates the exact numerical evaluation of
Eq.~\eqref{eq:DM} for small enough $\alpha\lesssim1$ as we show
in Fig.~\ref{fig:Asymptotic-decoherence-factor}.

On the other hand, in the limit of large $\alpha$, the coherent states
$|k\alpha\rangle_{E}$ become almost orthogonal for different $k$'s
and we can treat them approximately as the basis states. We can therefore
approximate the scalar product $D_{M}=|_{E}\langle C_{M}(\alpha)|U_{E}(\alpha)|C_{M}(\alpha)\rangle_{E}|$
by 
\begin{equation}
\begin{split}D_{M}(\alpha\to\infty) & \approx\frac{\sum_{k=0}^{M-1}{M \choose k}{M \choose k+1}}{\sum_{k=0}^{M}{M \choose k}{M \choose k}}=\frac{4^{M}\Gamma\left[M+\frac{1}{2}\right]}{\sqrt{\pi}M!{2M \choose M}}\\
 & =\frac{M}{M+1}=1-\frac{1}{M+1}
\end{split}
.\label{eq:DMinf}
\end{equation}
It has the form of Eq.~\eqref{eq:DMasympt} and very well approximates
the dephasing factor $D_{M}$ obtained numerically for large $\alpha\gtrsim4$
as also seen in Fig.~\ref{fig:Asymptotic-decoherence-factor}.

\subsection{Parameters used in the model QD Hamiltonian}

Exciton-phonon interaction constants $f_{\boldsymbol{k}}$ in the
deformation-potential coupling Hamiltonian $H_{{\rm int}}=|1\rangle\langle1|\sum_{\bm{k}}(f_{\bm{k}}^{*}b_{\bm{k}}+f_{\bm{k}}b_{\bm{k}}^{\dagger})$
are given by (in this part we reinsert $\hbar$ into the expressions)\begin{widetext}
\begin{equation}
f_{\bm{k}}=-i(\sigma_{\mathrm{e}}-\sigma_{\mathrm{h}})\sqrt{\frac{\hbar k}{2\varrho Vv}}\int_{\mathsf{\mathbb{R}^{3}}}d^{3}\bm{r}\psi^{*}(\bm{r})e^{-i\bm{k}\cdot\boldsymbol{r}}\psi(\bm{r})=-i(\sigma_{\mathrm{e}}-\sigma_{\mathrm{h}})\sqrt{\frac{\hbar k}{2\varrho Vv}}e^{-\frac{a^{2}(k_{x}^{2}+k_{y}^{2})+c^{2}k_{z}^{2}}{4}},
\end{equation}
where $\varrho=5360\mathrm{\, kg.m^{-3}}$ is the crystal density,
$V$ is the volume of the phonon system, $\sigma_{\mathrm{e/h}}$
($\sigma_{\mathrm{e}}=8\,\mathrm{eV},\,\sigma_{\mathrm{h}}=-1\,\mathrm{eV};\,\sigma_{\mathrm{e}}-\sigma_{\mathrm{h}}=9\,\mathrm{eV}$)
are deformation potential constants for electrons and holes, $v=5100\,\mathrm{m.s^{-1}}$
speed of longitudinal sound waves \cite{roszak10}, and $\psi(\bm{r})=\exp\left[-(x^{2}+y^{2})/2a^{2}-z^{2}/2c^{2}\right]/\sqrt[4]{\pi^{3}a^{4}c^{2}}$
are the exciton wave functions modeled by anisotropic Gaussians with
$a=5$ nm width in the $xy$-plane and $c=1$ nm along the $z$-axis.
Therefore, we get for the spectral density (recall that $\omega_{\boldsymbol{k}}=v|\boldsymbol{k}|$)
\begin{equation}
\begin{split}J(\omega>0) & =\frac{1}{\hbar^{2}}\sum_{\bm{k}}|f_{\bm{k}}|^{2}\delta(\omega-\omega_{\bm{k}})=\frac{(\sigma_{\mathrm{e}}-\sigma_{\mathrm{h}})^{2}}{2\hbar\varrho v}\frac{1}{V}\sum_{\bm{k}}ke^{-\frac{a^{2}(k_{x}^{2}+k_{y}^{2})+c^{2}k_{z}^{2}}{2}}\delta(\omega-\omega_{\bm{k}})\\
 & =\frac{(\sigma_{\mathrm{e}}-\sigma_{\mathrm{h}})^{2}}{2\hbar\varrho v}\frac{1}{(2\pi)^{3}}\int_{\mathsf{\mathbb{R}^{3}}}d^{3}\bm{k}ke^{-\frac{a^{2}(k_{x}^{2}+k_{y}^{2})+c^{2}k_{z}^{2}}{2}}\delta(\omega-vk)\\
 & =\frac{(\sigma_{\mathrm{e}}-\sigma_{\mathrm{h}})^{2}}{2\hbar\varrho v}\frac{1}{(2\pi)^{2}}\int_{0}^{\infty}dkk^{3}\delta(\omega-vk)\int_{0}^{\pi}d\theta\sin\theta e^{-\frac{k^{2}}{2}(a^{2}\sin^{2}\theta+c^{2}\cos^{2}\theta)}\\
 & =\frac{(\sigma_{\mathrm{e}}-\sigma_{\mathrm{h}})^{2}}{\hbar\varrho v^{5}(2\pi)^{2}}\times\omega^{3}e^{-\frac{\omega^{2}c^{2}}{2v^{2}}}\times\frac{1}{2}\int_{0}^{\pi}d\theta\sin\theta e^{-\frac{\omega^{2}c^{2}}{2v^{2}}\left(\frac{a^{2}}{c^{2}}-1\right)\sin^{2}\theta}\\
 & =\eta\omega^{3}e^{-\frac{\omega^{2}}{\omega_{c}^{2}}}F\left(\frac{\omega}{\omega_{c}}\right),
\end{split}
\end{equation}
\end{widetext} with the coefficient $\eta\equiv(\sigma_{\mathrm{e}}-\sigma_{\mathrm{h}})^{2}/\hbar\varrho v^{5}(2\pi)^{2}\doteq0.027\,\mathrm{ps^{2}}$,
cut-off frequency $\omega_{c}\equiv\sqrt{2}v/c\doteq7.21\,\mathrm{ps^{-1}}$
and the function $F(x)$ given by the last integral expression whose
asymptotic behavior for small and large $x$ is stated (for $a/c=5$)
in the main text.

\subsection{Derivation of Eqs. \eqref{densitymatrices} and \eqref{eq:D1}}

As mentioned in the main text the time evolution $\sigma(t)=e^{-iHt}\sigma(0)e^{iHt}$
of a factorizing initial state of the qubit plus the phonon environment
in the form $\sigma(0)=\rho(0)\otimes\varrho_{\mathrm{init}}$ can
be solved formally exactly by employing the Weyl operator $W(t)\equiv e^{iB(-t)}e^{-iB(0)}$
following the chain of arguments (recall that $H=S^{\dagger}H_{0}S$
with $S=\exp(-i|1\rangle\langle1|B)$ and $H_{0}=\epsilon|1\rangle\langle1|+H_{\mathrm{ph}}$)
\begin{equation}
\begin{split}\sigma(t) & =S^{\dagger}e^{-iH_{0}t}S\rho(0)\otimes\varrho_{\mathrm{init}}S^{\dagger}e^{iH_{0}t}S\\
 & =S^{\dagger}S(-t)e^{-iH_{0}t}\rho(0)\otimes\varrho_{\mathrm{init}}e^{iH_{0}t}S^{\dagger}(-t)S.
\end{split}
\end{equation}
From the definition of $S$ we get $S^{\dagger}(-t)S=|0\rangle\langle0|+|1\rangle\langle1|e^{iB(-t)}e^{-iB(0)}=|0\rangle\langle0|+|1\rangle\langle1|W(t)$
and, using the initial pure state of the qubit $|\mathrm{init}\rangle=(|0\rangle+e^{i\phi_{\mathrm{init}}}|1\rangle)/\sqrt{2}$,
we can write for $\sigma(t)$ in the block matrix form in the qubit
basis $\left\{ |0\rangle,|1\rangle\right\} $
\begin{equation}
\sigma(t)=\frac{1}{2}\begin{pmatrix}\varrho_{\mathrm{init}}(t) & e^{i(\epsilon t-\phi_{\mathrm{init}})}\varrho_{\mathrm{init}}W(t)\\
e^{-i(\epsilon t-\phi_{\mathrm{init}})}W^{\dagger}(t)\varrho_{\mathrm{init}}(t) & W^{\dagger}(t)\varrho_{\mathrm{init}}(t)W(t)
\end{pmatrix},\label{eq:sigma}
\end{equation}
with $\varrho_{\mathrm{init}}(t)=e^{-iH_{\mathrm{ph}}t}\varrho_{\mathrm{init}}e^{iH_{\mathrm{ph}}t}$.
The projective measurement onto states $|\pm\rangle$ at time $\tau$
then yields Eq.~\eqref{measurement} with the (normalized) phonon
bath density matrices $\varrho_{\pm}(\tau^{+})\propto\langle\pm|\sigma(\tau)|\pm\rangle$
stemming from Eq.~\eqref{eq:sigma} (with the help of relation $W(t)W^{\dagger}(t)=\mathbb{I}$)
and given in Eq.~\eqref{densitymatrices}.

We can use Eq.~\eqref{eq:sigma} also for the subsequent time evolution
of the density matrix for time $t$ after the measurement via replacing
$\varrho_{\mathrm{init}}(t)$ by $\varrho_{\pm}(t+\tau)\equiv e^{-iH_{\mathrm{ph}}t}\varrho_{\pm}(\tau^{+})e^{iH_{\mathrm{ph}}t}$
since the total state of the system plus phonon reservoir just after
the measurement \eqref{measurement} is of the factorized form assumed
in its derivation. Consequently, we obtain for the off-diagonal element
of the qubit density matrix $\rho_{01}^{\pm}(t+\tau)=\pm\langle W(t)\rangle_{\varrho_{\pm}(t+\tau)}e^{i(\epsilon t-\phi)}/2=e^{i\epsilon t}\langle W(t)\rangle_{\varrho_{\pm}(t+\tau)}\rho_{01}^{\pm}(\tau^{+})$
and the degree of decoherence is determined by the quantity\begin{widetext}
\begin{equation}
\langle W(t)\rangle_{\varrho_{\pm}(t+\tau)}=\mathrm{Tr_{ph}}\left(W(t)e^{-iH_{\mathrm{ph}}t}\varrho_{\pm}(\tau^{+})e^{iH_{\mathrm{ph}}t}\right)\equiv\mathrm{Tr_{ph}}\left(\widetilde{W}(t)\varrho_{\pm}(\tau^{+})\right),
\end{equation}
with $\widetilde{W}(t)\equiv e^{iH_{\mathrm{ph}}t}W(t)e^{-iH_{\mathrm{ph}}t}=e^{iB(0)}e^{-iB(t)}$.
Using the fact that the very initial state of the phonon reservoir
was canonical and, therefore, also $\varrho_{\mathrm{init}}(\tau)=\varrho_{\mathrm{ph}}^{\mathrm{can}}$
we can write (recall that $\left\langle \bullet\right\rangle _{0}\equiv\mathrm{Tr_{ph}}(\bullet\varrho_{\mathrm{ph}}^{\mathrm{can}})$)
\begin{equation}
\langle W(t)\rangle_{\varrho_{\pm}(t+\tau)}=\frac{\langle\widetilde{W}(t)\rangle_{0}+\langle W(\tau)\widetilde{W}(t)W^{\dagger}(\tau)\rangle_{0}\pm\left[e^{i\Delta\phi(\tau)}\langle W(\tau)\widetilde{W}(t)\rangle_{0}+e^{-i\Delta\phi(\tau)}\langle\widetilde{W}(t)W^{\dagger}(\tau)\rangle_{0}\right]}{2\left(1\pm\Re\left[e^{i\Delta\phi(\tau)}\langle W(\tau)\rangle_{0}\right]\right)}.
\end{equation}
The required mean values are calculated with the help of cumulants
(due to the Gaussian nature of the canonical density matrix the second
cumulants give \emph{exact} results --- see, e.g., Ref.~\cite[Sec.\ 4.3.2]{mahan00})
\begin{equation}
\begin{aligned}\langle\widetilde{W}(t)\rangle_{0} & =\langle W(t)\rangle_{0}=\exp\left[-w(t)\right],\\
\langle W(\tau)\widetilde{W}(t)\rangle_{0} & =\langle e^{iB(-\tau)}e^{-iB(0)}e^{iB(0)}e^{-iB(t)}\rangle_{0}=\langle e^{iB(-\tau)}e^{-iB(t)}\rangle_{0}=\langle W(t+\tau)\rangle_{0}=\exp\left[-w(t+\tau)\right],\\
\langle\widetilde{W}(t)W^{\dagger}(\tau)\rangle_{0} & =\langle e^{iB(0)}e^{-iB(t)}e^{iB(0)}e^{-iB(-\tau)}\rangle_{0}=\exp\left[-w(t)-w(-t)-2w(-\tau)+w(-t-\tau)\right],\\
\langle W(\tau)\widetilde{W}(t)W^{\dagger}(\tau)\rangle_{0} & =\langle e^{iB(-\tau)}e^{-iB(t)}e^{iB(0)}e^{-iB(-\tau)}\rangle_{0}=\exp\left[-w(-t)+w(\tau)-w(-\tau)-w(t+\tau)+w(-t-\tau)\right],
\end{aligned}
\end{equation}
which eventually yields Eq.~\eqref{eq:D1}.\end{widetext}

%

\end{document}